\documentstyle{article}

\begin{document}

\title{The Barut Second-Order Equation, Dynamical Invariants and Interactions}

\author{Valeri V. Dvoeglazov\\
\small{Universidad de Zacatecas}\\
\small{Apartado Postal 636, Suc. UAZ}\\
\small{Zacatecas 98062 Zac. M\'exico}\thanks{A talk given at the {\it VI Mexican School on Gravitation and Mathematical Physics ``Approaches to Quantum Gravity", Nov. 21-27, 2004,
Playa del Carmen, QR, M\'exico.}}}

\date{February 28, 2005}

\maketitle

\bigskip

\begin{abstract}
The second-order equation in the $(1/2,0)\oplus (0,1/2)$ representation of the Lorentz group
has been proposed by A. Barut in the beginning of the 70s, ref.~\cite{barut}. It permits to explain the mass splitting
of leptons $(e,\mu,\tau)$. Recently, the interest has grown to this model (see, for instance, the papers by 
S. Kruglov~\cite{krug} and J. P. Vigier {\it et al.}~\cite{vig}). We continue the research
deriving the equation from the first principles, finding the dynamical invariants for this model,
investigating the influence of the potential interactions.
\end{abstract}


The Barut main equation is
\begin{equation}
[i\gamma_\mu \partial_\mu -\alpha_2 {\partial_\mu \partial_\mu \over m} +\kappa] \Psi =0\,.
\end{equation}
\begin{itemize}
\item
It represents a theory with the conserved current that is linear in 15 generators of the 4-dimensional representation of the $O(4,2)$ group, $N_{ab}= {i\over 2} \gamma_a \gamma_b, \gamma_a = \{ \gamma_\mu, \gamma_5, i\}$.

\item
Instead of 4 solutions it has 8 solutions with the correct relativistic relation $E=\pm \sqrt{{\bf p}^2 + m_i^2}$. In fact, it describes states of different masses (the second one is $m_\mu = m_e (1+ {3\over 2\alpha})$, $\alpha$ is the fine structure constant), provided that a certain 
physical condition is imposed on the $\alpha_2$ parameter (the anomalous magentic moment should be equal to $4\alpha/3$).

\item
One can also generalize the formalism to include the third state, the $\tau$- lepton~[1b].

\item
Barut has indicated at the possibility of including $\gamma_5$ terms ( e.g., 
$\sim \gamma_5 \kappa^{'}$).

\end{itemize}

If we present the 4-spinor as $\Psi ({\bf p}) = column (\phi_R ({\bf p}))\quad \phi_L ({\bf p}))$ then  Ryder states~\cite{ryder} that $\phi_R ({\bf 0}) =\phi_L ({\bf 0})$. Similar argument has been given by Faustov~\cite{faust}: ``the matrix $B$ exists such that $B u_\lambda ({\bf 0}) = u_\lambda ({\bf 0})$, $B^2 = I$ for any $(2J+1)$- component function within the Lorentz invariant theories"\footnote{The latter statement is more general than the Ryder one, because it admits
$B=\pmatrix{0& e^{+i\alpha}\cr
e^{-i\alpha}&0\cr}$, so that $\phi_R ({\bf 0}) = e^{i\alpha}\phi_L ({\bf 0})$.}
The most general form of the relation in the $(1/2,0)\oplus (0,1/2)$ representation  has been given by Dvoeglazov~[7,4a]:
\begin{equation}
\phi_L^h ({\bf 0}) = a (-1)^{{1\over 2}-h} e^{i(\theta_1 +\theta_2)} \Theta_{1/2}
[\phi_L^{-h} ({\bf 0})]^\ast
+ b e^{2i\theta_h} \Xi_{1/2}^{-1} [\phi_L^{h} ({\bf 0})]^\ast\,,
\end{equation}
with
\begin{equation}
\Theta_{1/2}= \pmatrix{0&-1\cr
1&0\cr}=-i\sigma_2\,,\quad \Xi_{1/2} =\pmatrix{e^{i\varphi}&0\cr
0&e^{-i\varphi}}\,,
\end{equation}
$\Theta_J$ is the Wigner operator for spin $J=1/2$, $\varphi$ is  the azimuthal angle ${\bf p}\rightarrow 0$ of the spherical coordinate system.

Next, we use the Lorentz transformations:
\begin{equation}
\Lambda_{R,L}= \exp (\pm {\bf \sigma}\cdot {\bf \phi}/2),\,\,cosh\, \phi = E_p/m,\,\,
sinh\, \phi = {\vert {\bf p}\vert /m},\,\, \hat\phi = {\bf p}/\vert{\bf p}\vert.\,
\end{equation}
Applying the boosts and the relations between spinors in the rest frame, one can obtain:
\begin{eqnarray}
\phi_L^h ({\bf p}) &=& a {p_0 - {\bf \sigma}\cdot {\bf p}\over m} \phi_R^h ({\bf p})
+ b (-1)^{{1\over 2}+h} \Theta_{1/2}\Xi_{1/2}\phi_R^{-h} ({\bf p}),\,\\
\phi_R^h ({\bf p}) &=& a {p_0 + {\bf \sigma}\cdot {\bf p}\over m} \phi_L^h ({\bf p})
+ b (-1)^{{1\over 2}+h} \Theta_{1/2}\Xi_{1/2}\phi_L^{-h} ({\bf p}).
\end{eqnarray}
($\theta_1=\theta_2 =0$, $p_0=E_p= \sqrt{{\bf p}^2+m^2}$).
In the Dirac form we have:
\begin{equation}
[a {\hat p\over m} -1] u_h ({\bf p}) + ib (-1)^{{1\over 2}-h}\gamma^5 {\cal C} u_{-h}^{\ast} ({\bf p}) =0,\,\label{df}
\end{equation}
where
${\cal C}= \pmatrix{0&i\Theta_{1/2}\cr
-i\Theta_{1/2}&0\cr}$. In the QFT form we must introduce the creation/annihilation operators. Let $b_\downarrow = -ia_\uparrow$, $b_\uparrow = +ia_\downarrow$, then
\begin{equation}
[a {i\gamma^\mu\partial_\mu \over m} + b {\cal C}{\cal K} - 1] \Psi (x^\nu)=0\,.
\end{equation} 
If one applies the unitary transformation to the Majorana representation~\cite{dvoeg3}
\begin{equation}
{\cal U}= {1\over 2}\pmatrix{1-i\Theta_{1/2}&1+i\Theta_{1/2}\cr
-1-i\Theta_{1/2}&1-i\Theta_{1/2}\cr}\,,\,\, {\cal U} {\cal C}{\cal K}{\cal U}^{-1} = -{\cal K}\,,
\end{equation}
then $\gamma$-matrices become to be pure imaginary, and the equations are pure real.
\begin{eqnarray}
\left [ a {i\hat \partial\over m} -b-1 \right ] \Psi_1&=&0\,,\\
\left [ a {i\hat \partial\over m} +b-1 \right ] \Psi_2&=&0\,,
\end{eqnarray}
where $\Psi =\Psi_1 + i\Psi_2$. It appears as if the real and imaginary parts have different masses. Finally, for superpositions $\phi=\Psi_1 +\Psi_2$, $\chi=\Psi_1-\Psi_2$, multiplying by $b\neq 0$ we have:
\begin{equation}
[2a {i\gamma^\mu \partial_\mu \over m} + a^2 {\partial^\mu \partial_\mu \over m^2}+b^2 -1]{\phi (x^\nu)\over \chi (x^\nu)} =0\,,
\end{equation}
If we put $a/2m \rightarrow \alpha_2$, ${1-b^2 \over 2a} m\rightarrow \kappa$ we recover
the Barut equation.

How can we get the third lepton state? See the refs.~[1b,4b]:
\begin{equation}
M_\tau = M_\mu + {3\over 2} \alpha^{-1} n^4 M_e = M_e + {3\over 2} \alpha^{-1} 1^4 M_e + {3\over 2} \alpha^{-1} 2^4 M_e = 1786.08 \, MeV\,.
\end{equation}
The physical origin was claimed by Barut to be in the magnetic self-interaction of the electron
(the factor $n^4$ appears due to the Bohr-Sommerfeld rule for the charge moving in circular orbits in the field of a fixed magnetic dipole ${\bf \mu}$).
One can start from (\ref{df}), but , as opposed to the above-mentioned, one can write the coordinate-space equation in the form:
\begin{equation}
[a {i\gamma^\mu\partial_\mu \over m} +b_1 {\cal C}{\cal K} -1 ]\Psi (x^\nu)
+b_2 \gamma^5 {\cal C}{\cal K} \tilde \Psi (x^\nu) =0\,,
\end{equation}
with $\Psi^{MR}=\Psi_1 + i\Psi_2$, $\tilde \Psi^{MR}= \Psi_3 +i\Psi_4$.
As a result,
\begin{eqnarray}
(a {i\gamma^\mu \partial_\mu \over m} -1) \phi -b_1 \chi+ib_2\gamma^5\tilde\phi &=&0\,,\\
(a {i\gamma^\mu \partial_\mu \over m} -1) \chi -b_1 \phi-ib_2\gamma^5\tilde\chi &=&0\,.
\end{eqnarray}
The operator $\tilde \Psi$ may be linear-dependent on the states included in the $\Psi$. let us apply the most simple form $\Psi_1=-i\gamma^5 \Psi_4$, $\Psi_2 =+i\gamma^5 \Psi_3$. Then, one can recover the 3rd order Barut-like equation~[4b]:
\begin{equation}
[i\gamma^\mu \partial_\mu -m {1\pm b_1 \pm b_2 \over a}] [i\gamma^\nu \partial_\nu + {a\over 2m}\partial^\nu \partial_\nu + m {b_1^2-1\over 2a}]\Psi_{1,2}=0\,.
\end{equation}
Thus, we have three mass states.

Let us reveal the connections with other models. For instance, in refs.~\cite{vig,feyn}
the following equation has been studied:
\begin{equation}
[(i\hat\partial -e\hat A) (i\hat\partial -e\hat A) -m^2 ]\Psi =
[(i\partial_\mu -e A_\mu) (i\partial^\mu -e A^\mu) -{1\over 2}e\sigma^{\mu\nu}F_{\mu\nu}-m^2 ]\Psi = 0
\end{equation}
for the 4-component spinor $\Psi$. This is the Feynman-Gell-Mann equation. In the free case we have the Lagrangian (see Eq. (9) of ref.~[3c]):
\begin{equation}
{\cal L}_0 = (i\overline{\hat\partial \Psi}) (i\hat \partial \Psi ) -m^2 \bar\Psi \Psi\,.\label{gm}
\end{equation}
We can note:
\begin{itemize}
\item
The Barut equation is a sum of the Dirac equation and the Feynman-Gell-Mann equation.

\item
Recently, it was suggested to associate an analogue of Eq. (\ref{gm}) with the dark matter~\cite{dm}, provided that $\Psi$ is composed of the self/anti-self charge conjugate spinors, and it has the dimension  $[energy]^{1}$ in $c=\hbar=1$. The interaction Lagrangian is ${\cal L}^H \sim g\bar\Psi\Psi \phi^2$.

\item
The term $\sim \bar\Psi \sigma^{\mu\nu} \Psi F_{\mu\nu}$ will affect the photon propagation, and non-local terms will appear  in higher orders.

\item
However, it was shown in~[3b,c] that a) the Mott cross-section formula (which represents the Coulomb scattering up to the order $\sim e^2$) is still valid; b) the hydrogen spectrum is not much disturbed; if the electromagnetic field is weak  the corrections are small.

\item
The solutions are the eigenstates of $\gamma^5$ operator.

\item
In general, $J_0$ is not the positive-defined quantity, since the general solution 
$\Psi = a\Psi_+ + b\Psi_-$, where $[i\gamma^\mu \partial_\mu \pm m]\Psi_{\pm}=0$, see also~\cite{mark}.
\end{itemize}

The most general conserved current of the Barut-like theories is
\begin{equation}
J_\mu = \alpha_1 \gamma_\mu +\alpha_2 p_\mu +\alpha_3 \sigma_{\mu\nu} q^\nu\,.
\end{equation}
Let us try the Lagrangian:
\begin{equation}
{\cal L}={\cal L}_{Dirac}+{\cal L}_{add}\,,
\end{equation}
\begin{eqnarray}
{\cal L}_{Dirac}&=& \alpha_1 [\bar\Psi \gamma^\mu (\partial_\mu \Psi) - (\partial_\mu \bar\Psi)\gamma^\mu \Psi] -\alpha_4 \bar \Psi \Psi\,,\\
{\cal L}_{add} &=& \alpha_2 (\partial_\mu \bar\Psi)(\partial^\mu \Psi) +\alpha_3 \partial_\mu \bar\Psi \sigma^{\mu\nu}\partial_\nu \Psi\,.
\end{eqnarray}
Then, the equation follows:
\begin{equation}
[2\alpha_1 \gamma^\mu \partial_\mu -\alpha_2 \partial_\mu\partial^\mu -\alpha_4]\Psi=0\,,
\end{equation}
and its Dirac-conjugate:
\begin{equation}
\bar\Psi [2\alpha_1 \gamma^\mu \partial_\mu +\alpha_2 \partial_\mu\partial^\mu +\alpha_4]=0\,.
\end{equation}
The derivatives acts to the left in the second equation. Thus, we have the Dirac equation when $ \alpha_1={i\over 2}$, $\alpha_2=0$, and the Barut equation when $\alpha_2={1\over m}{2\alpha/3\over 1+4\alpha/3}$.

In the Euclidean metrics the dynamical invariants are
\begin{equation}
{\cal J}_\mu =-i \sum_i [{\partial{\cal L}\over \partial (\partial_\mu \Psi_i)} \Psi_i - \bar\Psi_i {\partial{\cal L}\over \partial (\partial_\mu \bar \Psi_i)} ]\,,
\end{equation}
\begin{equation}
{\cal T}_{\mu\nu}=- \sum_i [{\partial{\cal L}\over \partial (\partial_\mu \Psi_i)} \partial_\nu\Psi_i + \partial_\nu\bar\Psi_i {\partial{\cal L}\over \partial (\partial_\mu \bar \Psi_i)} ]+{\cal L}\delta_{\mu\nu}\,,
\end{equation}
\begin{equation}
{\cal S}_{\mu\nu,\lambda}=- i\sum_{ij} [{\partial{\cal L}\over \partial (\partial_\lambda \Psi_i)}N_{\mu\nu,ij}^\Psi \Psi_j + \bar\Psi_i N_{\mu\nu,ij}^{\bar\Psi}{\partial{\cal L}\over \partial (\partial_\lambda \bar \Psi_j)} ]\,.
\end{equation}
$N_{\mu\nu}^{\Psi,\bar\Psi}$ are the Lorentz group generators.

Then, the energy-momentum tensor is
\begin{eqnarray}
{\cal T}_{\mu\nu}&=& -\alpha_1 [ \bar \Psi \gamma_\mu \partial_\nu \Psi - \partial_\nu \bar\Psi \gamma_\mu\Psi ] -\alpha_2 [ \partial_\mu \bar\Psi \partial_\nu \Psi +\partial_\nu \bar\Psi \partial_\mu \Psi ] -\alpha_3 \left [ \partial_\alpha \bar\Psi \sigma_{\alpha\mu}\partial_\nu \Psi +\right.\nonumber\\
&+&\left. \partial_\nu \bar\Psi \sigma_{\mu\alpha}\partial_\alpha \Psi \right ]+
\left [ \alpha_1 (\bar\Psi \gamma_\mu\partial_\mu \Psi -\partial_\mu \bar\Psi \gamma_\mu \Psi) +\alpha_2 \partial_\alpha \bar\Psi \partial_\alpha \Psi +\alpha_3 \partial_\alpha \bar\Psi \sigma_{\alpha\beta}\partial_\beta \Psi +\right. \nonumber\\
&+&\left. \alpha_4 \bar \Psi \Psi \right ]\delta_{\mu\nu}\,.
\end{eqnarray}
Hence, the Hamiltonian $\hat {\cal H}=-iP_4 = - \int {\cal T}_{44} d^3 x$ is
\begin{equation}
\hat{\cal H} =\int \{ \alpha_1 [ \partial_i \bar\Psi \gamma_i \Psi - \bar\Psi \gamma_i \partial_i \Psi ] +\alpha_2 [ \partial_4 \bar\Psi \partial_4 \Psi - \partial_i\bar\Psi \partial_i \Psi ] -\alpha_3 [ \partial_i \bar\Psi \sigma_{ij} \partial_j \Psi ] -\alpha_4 \bar\Psi\Psi \} d^3 x\,.
\end{equation}
The 4-current is
\begin{equation}
{\cal J}_\mu = -i \{ 2\alpha_1 \bar\Psi \gamma_\mu \Psi +\alpha_2 [ (\partial_\mu \bar\Psi) \Psi -\bar \Psi (\partial_\mu \Psi ) ] +\alpha_3 [\partial_\alpha \bar\Psi \sigma_{\alpha\mu} \Psi -\bar \Psi \sigma_{\mu\alpha} \partial_\alpha \Psi ] \}\,.
\end{equation}
Hence, the charge operator $\hat {\cal Q} =-i\int {\cal J}_4 d^3 x$ is
\begin{equation}
\hat{\cal Q} = -\int \{ 2\alpha_1 \Psi^\dagger \Psi +\alpha_2 [ (\partial_4 \bar\Psi) \Psi -\bar \Psi (\partial_4 \Psi ) ] + \alpha_3 [\partial_i \bar\Psi \sigma_{i4} \Psi -\bar \Psi \sigma_{4i} \partial_i \Psi ] \} d^3 x \,.
\end{equation}
Finally, the spin tensor is
\begin{eqnarray}
{\cal S}_{\mu\nu,\lambda} &=& -{i\over 2} \left\{ \alpha_1 [ \bar\Psi \gamma_\lambda \sigma_{\mu\nu}\Psi +\bar\Psi \sigma_{\mu\nu} \gamma_\lambda \Psi ] 
+ \alpha_2 [ \partial_\lambda \bar\Psi \sigma_{\mu\nu} \Psi - \bar\Psi \sigma_{\mu\nu} \partial_\lambda \Psi ] + \right.\\
&&\left.+\alpha_3 [ \partial_\alpha \bar\Psi \sigma_{\alpha\lambda}\sigma_{\mu\nu}\Psi - \bar\Psi \sigma_{\mu\nu}\sigma_{\lambda\alpha}\partial_\alpha \Psi ] \right\} \,.\nonumber
\end{eqnarray}

In the quantum case the corresponding field operators are written:
\begin{eqnarray}
\Psi (x^mu) &=& \sum_h \int {d^3{\bf p}\over (2\pi)^3} [u_h ({\bf p})a_h ({\bf p}) e^{+ip\cdot x} +v_h ({\bf p}) b_h^\dagger ({\bf p})e^{-ip\cdot x}]\,,\\
\bar\Psi (x^mu) &=& \sum_h \int {d^3{\bf p}\over (2\pi)^3} [\bar u_h ({\bf p})a_h^\dagger ({\bf p}) e^{-ip\cdot x} + \bar v_h ({\bf p}) b_h ({\bf p})e^{+ip\cdot x}]\,.
\end{eqnarray}
The 4-spinor normalization is
\begin{equation}
\bar u_h u_{h'} =\delta_{hh'}\,,\quad \bar v_h v_{h'}=-\delta_{hh'}\,.
\end{equation}
The commutation relations are
\begin{eqnarray}
\left [ a_h ({\bf p}), a_{h^\prime}^\dagger ({\bf k}) \right ]_+ &=& (2\pi)^3 {m \over p_4} \delta^{(3)} ({\bf p}-{\bf k}) \delta_{hh^\prime}\,,\\
\left [ b_h ({\bf p}), b_{h^\prime}^\dagger ({\bf k}) \right ]_+ &=& (2\pi)^3 {m \over p_4} \delta^{(3)} ({\bf p}-{\bf k}) \delta_{hh^\prime}\,,
\end{eqnarray}
with all other to be equal to zero. The dimensions of the $\Psi$, $\bar\Psi$
are as usual, $[energy]^{3/2}$. Hence, the second-quantized Hamiltonian is written
\begin{equation}
\hat{\cal H} = - \sum_h \int {d^3{\bf p}\over (2\pi)^3} {2E_p^2 \over m} [\alpha_1 +m\alpha_2]
:[a_h^\dagger a_h - b_h b_h^\dagger]:\,.
\end{equation}
(Remember that $\alpha_1 \sim {i\over 2}$, the commutation relations may give another $i$, so the contribution of the first term to eigenvalues will be real. But if $\alpha_2$ is real, the contribution of the second term may be imaginary). The charge is
\begin{equation}
\hat {\cal Q} =- \sum_{hh'} \int {d^3{\bf p}\over (2\pi)^3} {2E_p \over m} [(\alpha_1 +m\alpha_2) \delta_{hh'} -i\alpha_3 \bar u_h \sigma_{i4} p_i u_{h'}] :[a_h^\dagger a_{h'}+b_h b_{h'}^\dagger]:\,.
\end{equation}
However, due to $[\Lambda_{R,L}, {\bf \sigma}\cdot {\bf p}]_- =0$ the last term with $\alpha_3$ does not contribute.

The conclusions are:

\begin{itemize}

\item
We obtained the Barut-like equations of the 2nd order and 3rd order in derivatives. The Majorana representation has been used.

\item
We obtained dynamical invariants for the free Barut field on the classical and quantum level.

\item
We found relations with other models (such as the Feynman-Gell-Mann equation).

\item
As a result of analysis of dynamical invariants, we can state that at the free level
the term $\sim \alpha_3 \partial_\mu \bar \Psi \sigma_{\mu\nu}\partial_\nu \Psi$ in
the Lagrangian does not contribute.

\item
However, the interaction terms $\sim \alpha_3 \bar\Psi \sigma_{\mu\nu}\partial_\nu \Psi A_\mu$
will contribute when we construct the Feynman diagrams and the $S$-matrix. In the curved space (the 4-momentum Lobachevsky space) the influence of such terms has been investigated in the Skachkov works~\cite{skach}. Briefly, the contribution will be such as if the 4-potential were interact with some ``renormalized" spin. Perhaps, this explains, why did Barut use the classical anomalous magnetic moment $g\sim 4\alpha /3$ instead of $\alpha \over 2\pi$.

\end{itemize}

The author acknowledges discussions with participants of recent conferences.

\end{document}